# Rightsizing LISA

R T Stebbins
NASA Goddard Space Flight Center, Code 663, Greenbelt, MD 20771 USA
Email: Robin.T.Stebbins@nasa.gov

**Abstract**. The LISA science requirements and conceptual design have been fairly stable for over a decade. In the interest of reducing costs, the LISA Project at NASA has looked for simplifications of the architecture, at downsizing of subsystems, and at descopes of the entire mission. This is a natural activity of the formulation phase, and one that is particularly timely in the current NASA budgetary context. There is, and will continue to be, enormous pressure for cost reduction from both ESA and NASA, reviewers and the broader research community. Here, the rationale for the baseline architecture is reviewed, and recent efforts to find simplifications and other reductions that might lead to savings are reported. A few possible simplifications have been found in the LISA baseline architecture. In the interest of exploring cost sensitivity, one moderate and one aggressive descope have been evaluated; the cost savings are modest and the loss of science is not.

## 1. Introduction

In the conceptual formulation of a flight mission, there is a constant tension between the desire to do more science or to add more functionality or reliability and the pressure to reduce complexity and costs. The science community wants the most science from a mission concept that can be selected and successfully flown. However, the 'selection process' is complex and difficult to anticipate. The goal of the project team is to formulate a robust, buildable concept for the construction, integration, testing and operational phases of the mission. There is - and will continue to be - pressure to reduce cost from agencies, reviewers, and the broader research community. There have been - and will be more - opportunities to augment the scope of LISA science. Exploring trade-offs between science and cost is central to mission formulation.

The Laser Interferometer Space Antenna (LISA) is a joint ESA-NASA mission to design, build and operate a spaceborne gravitational wave detector. The science requirements and the concept have been stable for over a decade [1]. The detector consists of a triangular constellation of three spacecraft in earth-like orbits separated by 5 million kilometers. Gravitational waves are detected by interferometrically measuring changes in the separation of free-falling proof masses. At each vertex of the triangle, a spacecraft encloses two proof masses defining the endpoints of the monitored separations.

At the request of NASA Headquarters, the LISA Project team in the U.S. has made a significant effort in the past year to search for cost savings and evaluate the trade-offs between science and cost. This paper reports the results of that search and presents a quantitative analysis of the trade-offs between science and cost.

Three large questions loom over any such assessment: Can the cost of the baseline design be reduced? What can be saved through descopes? Is the baseline design close to optimal? There are no definitive answers to these questions, only strong suggestions from the work done to date. And a clever innovation can always change the answers.

**2. The Search for Savings and the Approach to Descopes**

The LISA Project is in the so-called "formulation phase." During formulation, the science definition team develops and refines the science requirements that set the scope of the mission. At the same time, the project team refines and optimizes the conceptual design, and assesses both performance and programmatic requirements. Any design must have at least the minimal equipment to meet the science requirements. The architecture must also meet the redundancy and probability-of-success requirements levied by the agencies' engineering and risk policies, respectively.

The search for savings delved into several areas. We looked for architectural simplifications, such as the elimination of nonessential subsystems that had been included for extra robustness or desirable, but not essential, functionality. We looked for reductions of equipment or consumables, such as propellant. We looked for nonessential components within subsystems, parts and assemblies that could be made smaller, cheaper or more efficient. We sought mass reductions that would require a smaller launch vehicle, the single largest cost element. That naturally led to reductions in propellant for the ~13 month transfer phase to the final operational orbit.

We sought to reduce complexity by simplifying subsystems and combining avionics boxes that would reduce the number of deliverables and tests. We considered accepting higher levels of risk through reduced redundancy, using lower grade parts and reduced testing. Normally, NASA would require a mission of LISA's cost and importance to be 'single-fault tolerant,' that is, not subject to loss of the mission by a single 'credible' fault, that is, where 'credible' excludes highly reliable elements like launch vehicle and structure. An exception to this policy could be requested to reduce the redundancy to 'selective redundancy,' wherein only more risky elements are redundant.

Finally, we explored ways to streamline activities, such as concurrent or staggered engineering, building and testing of multiple units. The principal goal here was to compress the schedule and reduce life cycle costs.

In the approach to descopes below, the rationale of formulation given above is to some extent reversed: The project team examined perturbations to the baseline design and the science team assessed the loss of science – without passing judgement on whether the reduced science was acceptable, i.e., the mission would still be selected. So, concept changes were picked first, and then science performance was evaluated.

LISA has another constraint that limits the architectural changes that merit consideration. At the urging of both NASA and ESA Headquarters, a technology demonstration mission was instigated in 2001. LISA Pathfinder is led by ESA, and NASA has contributed an instrumentation package. The general design philosophy of ESA's LISA Technology Package was to design for LISA, rather than the somewhat relaxed requirements of Pathfinder's less benign environment. Consequently, the technology being tested on Pathfinder cannot realistically be substantially changed in a modified, or descoped, LISA. For this reason, and because there are no beneficial modifications known, that part of the LISA architecture – which constitutes most of the science payload – was left unchanged in these descope studies. In effect, it is too late to make a substantial savings there, because major development costs have already been incurred.

Over the last couple of years, the joint ESA/NASA Project Team has put considerable effort into the search for concept changes that would lead to lower costs. A couple of dozen alternatives have been examined, and none have been found to result in significant savings. EADS Astrium GmbH, the ESA Formulation contractor, has also explored many design variations. While some have been incorporated into the baseline for other reasons, none of them has produced known savings. In the effort starting in Fall 2007, the Project first did a "bottoms-up" cost estimate of the baseline concept for the first time. Previous costings were based on parametric models that rely on a database of previous missions with similar characteristics (e.g., mass, pointing accuracy, power, etc). In a "bottoms-up" cost estimate, one maps out the work elements to be performed at some level of refinement and estimates the cost of each one. At the element level, the cost may be based on either estimates of labor and procurements, the cost of that subsystem in a comparable mission or a parametric model for that subsystem. For example, a power subsystem might be based on the well-known costs of solar arrays, power conditioners, and the labor to assemble and test them. Flight software would be costed from a completed mission with comparable functionality. The bottoms-up cost model was informed by the advanced state of the LISA Pathfinder mission where appropriate. It was used for some of the cost estimates reported here, especially those involving the scientific payload, mission science and operations.

The cost estimate for the spacecraft, propulsion module, and launch vehicle were provided by the Mission Design Lab (MDL) of the Goddard Space Flight Center. The MDL is a rapid, concurrent design facility of a sort common in large spaceflight organizations. A team of experienced engineers representing all of the major disciplines (mechanical, thermal, power, avionics, propulsion, reliability, etc.) cooperate to develop and detail designs in a collaborative environment. Each discipline engineer designs and costs his subsystem using his experience, a database of relevant components, a suite of analysis software and coordinated costing tools. A systems engineer oversees the process to ensure a compatible overall design and consistent costing assumptions. The MDL team analyzed, detailed and costed the mature LISA baseline, and then examined two descope options. Their review of the baseline resulted in some cost savings described in the next section. The MDL applied consistent engineering discipline to all three concepts.

The MDL team did not assess the scientific payload or the micronewton thrusters, because of their specialized nature. The Project team relied on the previous technical studies and the bottoms-up cost estimate for these elements. As mentioned, these elements can not be changed substantially because of LISA Pathfinder, and their costs are informed by the LISA Pathfinder experience.

The science performance of the baseline and two descope concepts was evaluated by the LISA Parameter Estimation Task Force and described in this symposium by Curt Cutler [2]. The Task Force's methodology is reported elsewhere in these proceedings [3]. In brief, given an instrument performance, they use the Fisher Matrix method and astrophysically motivated source numbers to predict how many sources will be detected during nominal operations with relevant levels of performance. Depending on the source type, a relevant level of performance may be signal-to-noise ratio or luminosity distance error or sky localization, for example. In this manner, the science performance of the three concepts can be compared, and the price/performance trade can be quantified.

## 3. The Baseline

In this section, we describe the minimum science instrument necessary to satisfy the LISA science requirements, the minimum spacecraft and support equipment required to support the scientific instrument, the cost savings found in the baseline and the design parameters of the baseline concept that were modified for the descopes. Although the search for descopes went beyond simply scaling these parameters, only descopes with these parameters adjusted were judged viable. This is an implicit indication that the baseline concept is close to minimal; no significant architectural reduction was found.

To begin with, the following rationale leads to the minimum equipment needed for the science instrument: The basic LISA measurement concept calls for measuring changes of a path length defined by two free-falling fiducial points. A passing gravitational wave manifests itself as a strain in space-time, i.e., a fractional change in lengths. The LISA science requirements [4] recognize that the strongest astrophysical sources known require a strain sensitivity, expressed as an amplitude spectral density, of roughly $10^{-20}/\sqrt{Hz}$ in a frequency band around $10^{-3}$ Hz. Free-falling test masses are the fiducial points to which the measurement is made. These test masses must be protected from very small disturbances that would cause them to move in a manner that would mimic a gravitational wave. Laser interferometry is the only known measurement system capable of measuring changes of tens of picometers over distances of millions of kilometers, as required for the strain sensitivity. The frequency noise of available lasers, even after pre-stabilization, demands that changes in two nearly-equal arms be measured and combined so as to segregate the laser frequency noise into the sum of the path lengths and the quadrupolar gravitational wave signal into the difference. So, as a minimum, the science requirements mandate three test masses defining two measurement paths millions of kilometers long and the laser interferometry equipment to measure changes in those lengths to the requisite sensitivity. These two long measurement paths dictate at least three spacecraft.

Single-fault tolerance demands a fourth, backup test mass should one of the others fail. But this fourth test mass must be positionable to any of the three spacecraft. Having two test masses per spacecraft is cheaper than a fourth spacecraft with adequate propulsion to get it to the required location. Arranging three spacecraft with two test masses each in an approximate equilateral triangle, and measuring all three sides satisfies this requirement with the minimum additional equipment. It has the additional benefits that the three spacecraft can be essentially identical, and, so long as all three paths are still being measured, two polarizations of the gravitational waves are simultaneously being measured. Having 6 proof masses active is easier than 3 active and 3 in stand-by. Hence the architecture of three spacecraft, each with two test masses at each vertex of an equilateral triangle defining the endpoints of two measurement paths, is the most efficient way to achieve redundancy.

The disturbance requirement, naturally expressed as an amplitude spectral density of residual acceleration, demands "drag-free" station-keeping. By sensing the location of the test mass with respect to the spacecraft, and forcing the spacecraft to follow the test mass by means of micronewton thrusters, unwanted force gradients associated with the spacecraft are not converted into time-varying forces on the test masses that could be mistaken for gravitational waves. Hence the low disturbance requirement leads to capacitive sensing and actuation, charge control, caging and a vacuum system for the test masses and micronewton thrusters for spacecraft station-keeping.

The precision metrology function requires the usual ingredients of laser interferometry: a laser source, frequency and power pre-stabilization, a transmit/receive telescope for beam expansion/compression, beam steering and combining optics, photoreceivers and a high resolution phasemeter. The latter is a reflection that LISA is a continuous ranging system that must resolve roughly $10^{-5}$ cycles/$\sqrt{Hz}$ to achieve 10 picometer sensitivity. Thus are the major constituents of the science instrument derived from the science requirements.

The LISA science instrument makes minimal demands on the spacecraft. None of the conventional spacecraft subsystems can be eliminated, but they require modest amounts of standard equipment. The opportunities for economy in the spacecraft lie in reduction of components and integration of subsystems.

To passively preserve the equilateral-triangle geometry of the measured paths requires heliocentric orbits with a fixed relationship between eccentricity and inclination for all spacecraft, and a different phasing of the line of nodes for each. The distance from Earth sets the useful lifetime of the mission: the closer, the less time before the triangle distorts and becomes unusable. The 5 million kilometer path length and 50 million kilometer distance from the Earth have been chosen to optimize the sensitivity to known gravitational wave sources and to enable science operations for a nominal 5 years, extendable to 8.5, respectively. The length of the arms and the distance of the triangle from the Earth are the main parameters scaled in the descope concepts.

A separate propulsion module is necessary to transfer each spacecraft and insert it into its operational orbit. These modules comprise propulsion systems, tankage, propellant and structure. They are discarded with separation mechanisms after insertion of the spacecraft in their operational orbits.

LISA is not a pointed instrument, and has unusually simple operating modes. The ground system has to downlink data every second day for eight hours, except a few times per year when more frequent downloads are needed to support a massive black hole binary merger. There is an acquisition sequence where the three spacecraft have to point their lasers at each other, and lock the laser frequencies together. Science operations will involve monitoring data quality and processing the science data to identify sources and estimate source parameters. The data processing is substantial, but the final data volume is modest (~100 GB).

The complete description of the baseline LISA architecture is too extensive to be repeated here. It can be found in [5].

The initial task for the MDL Team was to identify cost savings in the baseline architecture. They identified three significant changes, and one virtual one. First, the integration of the power management into the Computer and Data Handling subsystem eliminated a separate avionics box and added interface cards and software to the onboard computer. Second, rate gyros were eliminated as nonessential. The high performance star trackers needed for acquisition were deemed adequate for attitude control when not in science modes. Third, the implementation schedule was compressed by a year to reduce workforce costs. A lower quote from NASA Launch Services for the Atlas V 531 launch vehicle produced the virtual savings; neither the architecture, nor the schedule, was changed. These savings are carried over into the descope concepts presented in the following sections, and do not appear in the cost comparisons between the baseline and descope concepts. Savings from lower class parts were identified, but deemed not compatible with governing policies. Those savings are excluded from the baseline descope cost estimates.

## 4. The Descopes

After 15 years of study by the community, the combined project teams and industry contractors, the baseline concept is the only one known that can detect gravitational waves with high certainty. Given that, descopes can only be obtained by scaling (1) the length of the measured paths, (2) the length of science operations and (3) the instrument sensitivity. The path length affects the usable bandwidth and the sensitivity. The length of science operations affects operations cost, component service life and the distance that the constellation needs to be from Earth.

Instrument sensitivity is most naturally expressed in terms of strain amplitude spectral density. Strain is a fractional length change, and here a displacement noise divided by the path length. At frequencies below a few millihertz, the instrument sensitivity is set by residual acceleration. The test mass systems flying on LISA Pathfinder are designed for the baseline sensitivity, and even order-of-magnitude sensitivity reductions would not produce cost savings. At higher frequencies, instrument sensitivity is set by the position noise of the interferometer. Shot noise is responsible for roughly half of that error budget. Hence, telescope size and laser power can be adjusted. However, the economics of those technologies favors adjusting the telescope size. By virtue of the strain sensitivity being a displacement noise – whether from residual accelerations or measurement noise – divided by the path length, the sensitivity over the usable bandwidth is affected by all three variables.

The choice of these parameters for the baseline concept and two descope concepts is shown in table 1. These values were chosen to explore the interplay between science performance and mission cost. Note that the sensitivity choices for the position noise from the interferometer are given in dimensionless strain to reflect the choice of path length as well. The acceleration noise is defined from 0.1 to 8 mHz, and the position noise is defined from 2 to100 mHz. A more extensive discussion of LISA requirements can be found in [6].

**Table 1.** Parameter values for the three mission concepts.

| Concept | Path Length | Sensitivity (Position/Path Length) | Sensitivity (Acceleration) | Science Operations Lifetime |
|---|---|---|---|---|
| Baseline | 5 Gm | $3.6\text{x}10^{-21}$ /√Hz | $3\text{x}10^{-15}$ m/s²/√Hz | 5 yrs |
| Descope I | 2.5 Gm | $7.2\text{x}10^{-21}$ /√Hz | $3\text{x}10^{-15}$ m/s²/√Hz | 3 yrs |
| Descope II | 1 Gm | $11\text{x}10^{-21}$ /√Hz | $3\text{x}10^{-15}$ m/s²/√Hz | 2 yrs |

The relative science performance of the baseline concept and the two descopes is shown in table 2. The table shows the number of gravitational wave sources of three types that meet criteria representative of the LISA science objectives [4].

**Table 2.** Relative science performance for the three mission concepts.

| Source Type | Criterion | Baseline | Descope I | Descope II |
|---|---|---|---|---|
| Massive black hole mergers | Mergers with SNR>10 | 75-150 | 45-65 | 25-35 |
| | Mergers with distance uncertainty of <10% | 21 | 6 | 2 |
| | Mergers with sky localization <10 deg² | 5 | 1.5 | 0.5 |
| Extreme Mass Ratio Inspirals | Events with SNR>30 | 260 | 50 | 1 |
| Ultra-Compact Galactic Binaries | Binaries with SNR>10 | 17,000 | 7,000 | 2,500 |
| | Binaries with both distance and sky location determined | 2,800 | 600 | 200 |

A full discussion of LISA science [7] is necessary to appreciate the numbers in table 2, but a summary of the consequences will be indicative. To start, the mergers of massive black hole binaries are expected to happen as a result of the hierarchical mergers of galaxies and proto-galaxies since the end of the cosmological Dark Ages. Observations of a large enough number of these sources with sufficient signal-to-noise ratio (SNR) will provide extraordinary information about the formation, growth and merger history of massive black holes and their host galaxies. Standard LISA data processing will produce a source catalog with astrophysical parameters (e.g., luminosity distances, masses, spin vectors, orbital parameters, sky position, merger time) for each massive black hole binary.

The parameter uncertainty depends in a complicated way on the SNR, but generally uncertainties of all parameters increase with declining SNR. Luminosity distance is often the most poorly determined parameter, expressed fractionally. However, it is also one of the most valuable, since these luminosity distances will be the best direct distance measurements in astrophysics, and they will likely extend – with increasing uncertainty - to redshifts $z{\sim}10$. The ability to understand black hole formation and evolution requires knowledge of the distances to better than ± 10% (± 1σ). Descope II is unable to make meaningful statements about formation and evolution. Uncertainty in predictions of the underlying merger rates could reduce these numbers by another factor of two to four.

Sky localization could be very valuable if an electromagnetic counterpart to the merger event could be observed. If enough host galaxies can be identified and their redshifts obtained, then Dark Energy could be mapped to a few percent, with lower systematic errors than other methods. The 10 $deg^2$ error box corresponds to the anticipated field of view of the Large Synoptic Survey Telescope (LSST). Sky localization is at risk with Descope I, and lost for Descope II. Again, the numbers could be lower by a factor of two to four.

Extreme Mass Ratio Inspiral (EMRI) events involve the inspiral of white dwarves and stellar-mass neutron stars and black holes into massive black holes in galactic nuclei. EMRIs can be used to measure the mass and spin of massive black holes in the $z<1$ universe, to census the compact object population in galactic nuclei, to precisely test General Relativity in the strong field limit, and possibly to determine the Hubble constant. EMRI science is seriously degraded for Descope I and essentially lost for Descope II. EMRI event rates could be an order of magnitude lower, putting the science at risk even for Descope I.

Ultra-compact galactic binaries are combinations of white dwarfs, neutron stars and stellar mass black holes that have evolved to a very tight binary. A survey of these binaries will elucidate their formation, evolution and distribution throughout the Milky Way and environs. Their numbers will improve our understanding of white dwarfs, their masses, and their interactions. These sources are present in numbers so large that they will form a confusion threshold, and LISA will only be able to identify the most energetic individuals. All three concepts can detect significant numbers of binaries, but with differing capability of determining scientific characteristics, as illustrated by the numbers for which both distance and sky location can be determined. Scientific return for ultra-compact binaries is significantly enhanced by the ability to determine the spatial distribution of binaries, particularly for the less common subclasses.

In summary, all three concepts will be able to detect massive black hole binary mergers, but with sharply decreasing ability to measure scientifically important characteristics as instrument performance is descoped. Compared to the baseline, Descope I suffers: (1) a factor of 3 reduction in the number of massive black hole binaries which can be used for electromagnetic counterparts, (2) a reduction in the ability to test General Relativity in the strong field dynamical regime through a loss of SNR, (3) a loss of science and a factor of 5 reduction in EMRIs, (4) some loss of science with ultra-compact binaries, and (5) a loss of robustness against uncertainty in the underlying astrophysical rates. Descope II suffers: (1) loss of enough sources with good distance determination to study black hole formation and evolution, (2) an inability to determine electromagnetic counterparts for any massive black hole mergers, (3) loss of all EMRI science, (4) an order of magnitude loss of ultra-compact binary science, and (5) a pervasive loss of robustness against uncertainty in the underlying astrophysical rates.

The three concepts range from extraordinarily compelling and robust science performance to a much more circumscribed and risky performance for the range of instrument performance explored. Of the science from the baseline concept, the Beyond Einstein Program Assessment Review of the National Research Council said "On purely scientific grounds LISA is the mission that is most promising and least scientifically risky … Thus, the committee gave LISA its highest scientific ranking."[8]

**5. Concept Changes**

After scrubbing the baseline architecture for cost savings described in §3 above, the Goddard Mission Design Lab applied the Descope I parameters in table 2 to develop a Descope I concept and evaluate the cost savings. After a relatively thorough study of the Descope I concept wherein changes were made to most of the mission elements, a minimal effort was needed to change the concept and cost for the Descope II parameter values.

The major concept changes are shown in table 3. The Descope I concept incorporates several changes in the payload, spacecraft, propulsion module and launch vehicle:

- The laser power was reduced for the shorter arms.
- The telescope diameter was reduced for reduced sensitivity.
- Redundancy is selectively eliminated for propulsion module control electronics, solar array, star tracker, propulsion module thrusters and high gain antenna.
- The large apogee engine is eliminated from the propulsion module in favor of long firings by the small attitude thrusters to achieve the smaller ΔV (i.e., velocity changes in orbital maneuvers during transfer phase).
- Ka-band communications downlink eliminated in favor of higher power, bidirectional X-band with single larger high gain antenna.
- The launch mass is reduced by 1095 kg, mostly in propellant savings.
- Launch vehicle is reduced from Atlas V (531) to Atlas V (431).

The shorter measurement path length between spacecraft reduces propellant for the final injection maneuver. It also reduces the Earth's perturbation of the triangle; the constellation of spacecraft can then be placed closer to the Earth. Similarly, the shorter scientific operation permits moving the constellation still closer to the Earth since the Earth's perturbation of the equilateral geometry operates over a shorter period. The shorter path length also allows a reduction in laser power at the same sensitivity. The closer proximity to Earth reduces the demand for mission ΔV and saves significant propellant mass, which in turn allows a smaller launch vehicle.

**Table 3.** Design comparison between the three mission concepts.

| | **Baseline** | **Descope I** | **Descope II** |
|---|---|---|---|
| **Path Length** | $5\times10^9$ m | $2.5\times10^9$ m | $1\times10^9$ m |
| **Orbit** | 22° behind Earth | 12° behind Earth | 9° behind Earth |
| **Mission lifetime** | 6.5 yrs nominal; 10 yrs goal | 4.5 yrs nominal; 6.5 yrs goal | 2 yrs nominal |
| **Redundancy** | Single-fault tolerant design | Selective redundancy | Selective redundancy |
| **Mission ΔV** | 1130 m/s | 517 m/s | 333 m/s |
| **Launch vehicle $C3(km^2/sec^2)$: 0.5** | Atlas V (531) 5165 kg | Atlas V (431) ~ 4065 kg | Atlas V (411) ~ 3814 kg |

Some of the sought after cost savings were either not found, or not taken. Specifically, no architectural simplification was found, the smaller telescope did not change the overall payload dimensions, single-string (i.e., no redundancy) was not compatible with NASA Policy Regulations despite the shorter mission life, and Class C parts were judged to have an unacceptable reliability.

The Descope II concept differs from Descope I by placing the constellation still closer to Earth and shortening the mission lifetime significantly. Note that proximity to Earth precludes extending the mission lifetime later, since distortion of the triangle by the Earth makes the instrument unusable after two years. This maximal descope reduces costs through further reduction in propellant leading to a still smaller launch vehicle and shorter operations.

The science-to-cost relationships of both descope concepts relative to the baseline concept are shown in figure 1. Where there is a simple metric, the Descope I value appears above the Descope II value in each element.

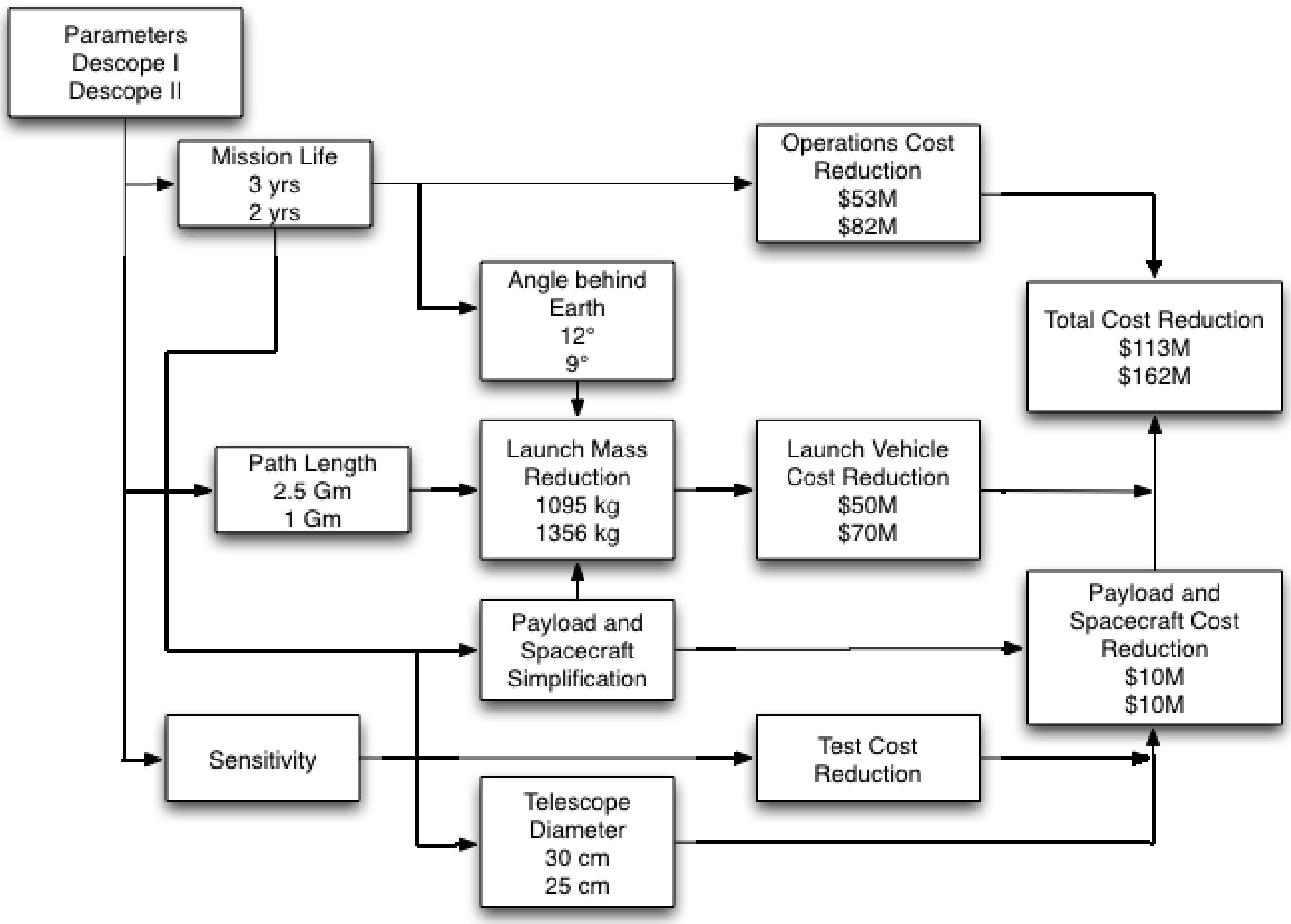

**Figure 1.** Science-to-cost relationships. In each box, the Descope I value appears above the Descope II numbers.

When compared against an approximately $1.2B total baseline concept cost (real year dollars, launch in 2018), the savings for both descope concepts are rather modest. The major sources of savings are the use of a smaller launch vehicle due to reduction in the required propellant and the lower operating cost for the reduced mission life time.

**6. Summary and Conclusions**

The U.S. science community and the U.S. Project team have analyzed the baseline LISA concept and two descope concepts to assess the science-to-cost relationships. Mission complexity and cost change very little with changes in science capability, even drastic reductions in science capability. Giving up more than half the science saves about 10% of the cost. Mission concepts with increased science scope were not studied, but past experience indicates that increases, say, in usable bandwidth, have substantial cost impact. This suggests that the LISA baseline concept is close to the "right" size.

There do not seem to be any substantial additional savings to be had in the spacecraft or operations. A radical idea that confronts complexity and/or redundancy in the payload seems unlikely.

**Acknowledgements**

Many members of the LISA community contributed to the results reported here. Peter Bender (Colorado) suggested the original descope levels. Stephen Merkowitz (GSFC) and Jeffrey Livas (GSFC) guided the studies at the Mission Design Lab. The LISA Parameter Estimation Task Force carried out science performance analyses, with a major contribution by Curt Cutler (JPL). Marta Volonteri (Michigan) generated the MBHB source distributions. Neil Cornish (Montana State) and Scott Hughes (MIT) supplied the parameter estimation code for the massive black hole binaries. Jonathon Gair (Cambridge) and Neil Cornish made important contributions to the EMRI evaluation. Thomas Prince (Caltech) summarized the relative science performance of the three concepts.